\documentclass[preprint,english,aps,showpacs]{revtex4-1}

\usepackage{amsmath}
\usepackage{color}
\usepackage{graphicx}

\usepackage{babel}

\begin{document}

\title{Residual disorder and diffusion in thin Heusler alloy films}
\author{B.~Jenichen}
\email{bernd.jenichen@pdi-berlin.de}
\author{J.~Herfort}
\author{T.~Hentschel}
\author{A.~Nikulin}
\email{on leave from Monash University, Clayton, Victoria 3800, Australia}
\author{X.~Kong}
\author{A.~Trampert}
\affiliation{Paul-Drude-Institut f\"ur Festk\"orperelektronik,
Hausvogteiplatz 5--7,D-10117 Berlin, Germany}
\author{I.~\v{Z}i\v{z}ak}
\affiliation{Institut f\"ur Nanometeroptik im Helmholtz-Zentrum fuer Materialien und Energie,
Albert-Einstein-Strasse 15, D-12489 Berlin, Germany}

\date{\today}

\begin{abstract}
Co$_{2}$FeSi/GaAs(110) and Co$_{2}$FeSi/GaAs(111)B hybrid structures were grown by
molecular-beam epitaxy and characterized by transmission electron
microscopy (TEM) and X-ray diffraction.  The films contained inhomogeneous distributions of
ordered $L2_1$ and $B2$ phases. The
average stoichiometry  was controlled by lattice parameter measurements, however
diffusion processes lead to inhomogeneities of the atomic concentrations and the degradation
of the interface, influencing long-range order. An average
long-range order of 30-60$\%$ was measured by grazing-incidence X-ray diffraction, i.e. the as-grown Co$_{2}$FeSi
films were highly but not fully ordered. Lateral inhomogeneities of the spatial distribution of long-range order in Co$_{2}$FeSi
were found using dark-field TEM images taken with
superlattice reflections.
\end{abstract}


\maketitle

\section{Introduction}
Device concepts based on the spin rather than the charge of the electron have been recently introduced in the field of spintronics.
These concepts are expected to lead to further improvements in device performance.
Heusler alloys can be useful for sources of spin injection into semiconductors
\cite{Palmstrom2003,zutic04,Felser09} as a first step for the fabrication of spintronic devices.~\cite{marrows2011}
The Heusler alloy Co$_{2}$FeSi (cubic $L2_1$ ordered structure, space group F$\bar{4}$3m, structure number 216 in Ref.~\onlinecite{hahn2006})
has some outstanding properties: It is a ferromagnetic half-metal with a Curie
temperature larger than 1100~K \cite{wurmehl05} and a magnetic moment of 6$\mu_B$.~\cite{wurmehl2006} The
lattice parameter (0.5658~nm) matches that of GaAs (0.5653~nm) \cite{hashimoto05}, i.e. it can be
grown epitaxially on GaAs in the thickness range of interest without the formation of misfit dislocations.
Therefore, Co$_{2}$FeSi is a promising material for spin injection into GaAs- or Ge-based structures
such as for example spin light-emitting diodes
\cite{Ohno99,zhu01,ramsteiner08}, magnetic tunnel junctions \cite{marukame2006,ishikawa2008, oogane2011} and spin field effect
transistors \cite{Sugahara04,Ando2010,Ando2011}. However interface disorder due to interdiffusion or chemical reaction
may be detrimental for technological application.~\cite{Gusenbauer2011,ramsteiner08}
X-ray and electron diffraction experiments yield information about
structure and long-range order of Heusler alloys.~\cite{webster1971,ziebeck1974,Jenichen05,takamura09,Jenichen2010}
In this study Co$_{2}$FeSi/GaAs(110) and Co$_{2}$FeSi/GaAs(111)B hybrid structures
grown by molecular-beam epitaxy (MBE) were investigated by transmission electron microscopy (TEM) and X-ray
diffraction (XRD) in order to characterize the stability of the ferromagnet/semiconductor
(FM/SC) interface and the structural properties of the Co$_{2}$FeSi film. A complete long-range ordering and a perfect
FM/SC interface seem to be inevitable for the utilization of the half-metallic properties of Co$_{2}$FeSi~\cite{miura2004},
i.e. to reach a high degree of spin polarization and successful spin injection into the semiconductor without severe scattering.
Significant diffusion of any of the constituents can hamper this goal. The GaAs(110) substrate
orientation has two big advantages: First, it has a longer spin lifetime compared to GaAs(001).~\cite{Ohno1999prl}
And second, the corresponding FM/SC interface is expected to
maintain the half-metallic properties~\cite{nagao2006}, similar as the (111) oriented interface.~\cite{Attema06}  The GaAs(111) substrate
orientation provides the FM/SC interface with the highest thermal stability.~\cite{Jenichen2010}

\section{Experiment}

Co$_{2}$FeSi films were grown on GaAs(110) and on GaAs(111)B substrates by
MBE as given in Refs.~(\onlinecite{Hentschel2012,Jenichen2010}).
The calibration of the fluxes was described in Ref.~(\onlinecite{hashimoto05}).
The growth rate was 0.1~nm~min$^{-1}$. The substrate temperature during MBE  growth
$T_{S}$ was varied between 100 and 350~$^{\circ}$C. The nominal Co$_{2}$FeSi film
thicknesses were 40~nm on GaAs(110) and 15~nm on GaAs(111)B. No additional capping layer was grown
on top of the Co$_{2}$FeSi.
The samples were investigated by dark-field and
high-resolution (HR) TEM. For that purpose cross-sectional TEM
specimens were prepared by mechanical lapping and polishing,
followed by argon ion milling according to standard techniques.
TEM images were acquired with a JEOL 3010 microscope operating at
300 kV. Electron energy loss spectroscopy (EELS) measurements were performed in the TEM
with a spot size of $\approx$10~nm. The Co/Fe ratio was determined by the analysis of
the Fe-L$_{2,3}$ and Co-L$_{2,3}$ edges after background subtraction.~\cite{egerton1996}
Since the analysis was carried out without standards only the lateral changes of the
composition ratio were determined in line scans. The cross section TEM methods provided
high lateral and depth resolutions on the nanometer scale, however they averaged
over the thickness of the thin sample foil ($\sim$~20~nm).
High-resolution XRD and X-ray reflectivity (XRR) measurements were performed
using a Panalytical X-Pert PRO MRD\texttrademark\ system
with a Ge(220) hybrid monochromator  (CuK$\alpha_1$ radiation with a
wavelength of $\lambda=1.54056$~\AA, spot size several mm$^2$). The simulation of X-ray reflectivity curves was
performed with the program Reflectivity\texttrademark\ provided by Panalytical. XRD patterns were calculated
in dynamical approximation \cite{Stepanov1997}. For the direct determination
of the displacement depth profile from the XRD curves we used the
X-ray phase retrieval method.~\cite{nikulin1998,nikulin1999}
We estimated the average long-range order using a comparison of the integrated intensities
of superlattice and fundamental reflections \cite{Jenichen05} measured using grazing
incidence diffraction (GID) of X-rays. We measured the 111, 222, and 220 reflections
with synchrotron radiation (energy 6900~eV, wavelength 0.179687~nm) at the beamline KMC2
of the electron storage ring BESSY II of the Helmholtz-Zentrum Berlin.

\section{Results and Discussion}

\begin{table}
\caption{Degree of average long-range order of Co$_{2}$FeSi films grown on a GaAs(110) substrate at two different temperatures} \vspace{12pt} \label{tab:data}
\begin{tabular}{ccccc}
~~~$T_S$~~~ &  ~~~S$_{B2}$~~~ & ~~~S$_{L21}$~~~& ~~error~~   \\
($^{\circ }$C)& (\%) & (\%)& (\%)  \\
\hline
100 &  32 & 48 & $\pm$2  \\
200 & 61 & 65 & $\pm$2  \\
\end{tabular}
\end{table}

\begin{figure}[!t]
\includegraphics[width=8.0cm]{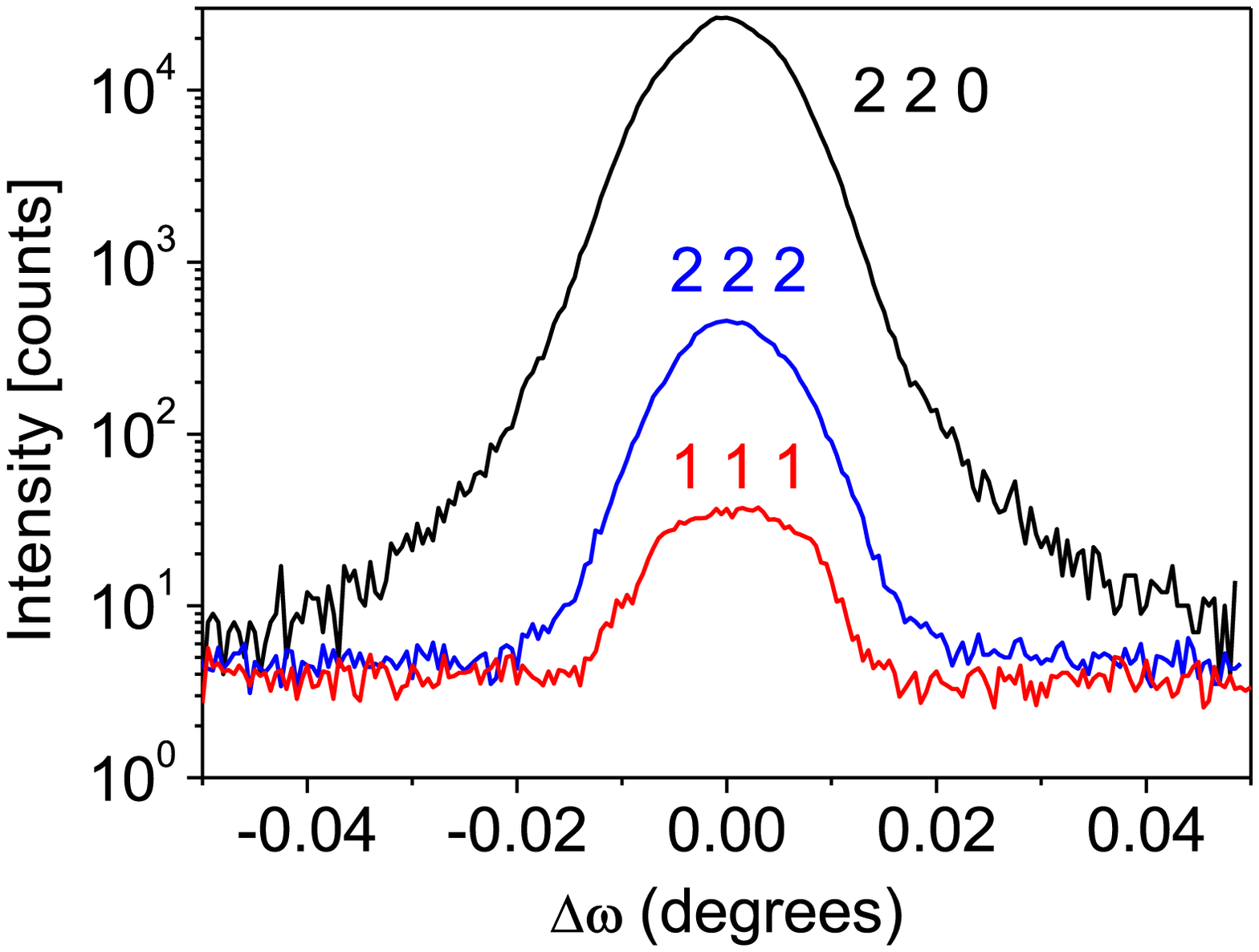}
\caption{(color online) Grazing incidence diffraction peaks of Co$_{2}$FeSi films of a nominal thickness of 40~nm on GaAs(110) at a substrate temperature $T_S$~=~200~$^{\circ}$C. The measurements are performed at an incidence angle $\alpha_i$~=~0.33$^{\circ}$ below the critical incidence angle of Co$_{2}$FeSi, $\alpha_i^{crit}$~=~0.426$^{\circ}$. Therefore the information depth is below 10~nm \cite{feidenhansl:review}, i.e. all the radiation is diffracted inside the 40~nm thick film. The fundamental 220 reflection and the 222 and 111 superlattice reflections are given.}
\label{fig:e6900compare}
\end{figure}

A perfectly ordered Co$_{2}$FeSi lattice is highly desirable to benefit
from the extraordinary properties of this half-metallic Heusler alloy. Interdiffusion
near the FM/SC interface can lead to the reduction of the order in the Co$_{2}$FeSi.
In order to determine the long-range order we can distinguish different types of diffraction peaks:
 The 220 reflection is fundamental (i.e. not sensitive to disorder)
whereas the 222 and the 111 reflections are superlattice reflections.~\cite{takamura09,Jenichen2010}
The 222 reflection arises when at least the CsCl-type $B2$ order is present in the Co$_{2}$FeSi lattice whereas the 111 reflection can be found only
in regions of $L2_1$ order. In the 110 oriented samples we
found all those reflections well oriented for GID measurements, i.e. the
diffracting netplanes are perpendicular to the surface. GID has the advantage of a
limited information depth for incidence (and/or exit) angles below the critical angle.~\cite{dosch86, feidenhansl:review}  In our case,
this information depth is smaller than the film thickness, i.e. using GID we measure the region of the film near the surface and exclude an influence of the substrate. Figure~\ref{fig:e6900compare} displays the three different types of GID peaks ($\omega/2\Theta$-scans) of a 40~nm thick Co$_{2}$FeSi film grown on  GaAs(110) at a substrate temperature $T_S$~=~200~$^{\circ}$C.

The degree of $B2$ ordering, $S_{B2}$, can be defined as follows:
\begin{equation}\label{equ1}
\begin{split}
S_{B2} = \frac{n_{Co}-n_{Co}^{random}} {n_{Co}^{full-order}-n_{Co}^{random}},
\end{split}
\end{equation}
where n$_{Co}^{random}$ is the number of Co atoms on Co sites for the most random distribution, i.e. the $A2$ structure, and
n$_{Co}^{full-order}$ is the number of Co atoms on Co sites in the ordered $B2$ structure.
$S_{B2}$ can be determined from XRD measurements using the relation between
the even superlattice and the fundamental reflections \cite{webster1971, ziebeck1974}:

\begin{equation}\label{equ2}
\begin{split}
\frac{I_{222}^{exp}} {I_{220}^{exp}} = S_{B2}^2~\frac{I_{222}^{full-order}} {I_{220}^{full-order}},
\end{split}
\end{equation}
where $I_{222}^{exp}$ and $I_{220}^{exp}$ are the measured integral intensities and $I_{222}^{full-order}$ and $I_{220}^{full-order}$ are calculated from the structure factors.~\cite{Jenichen2010}
The degree of $L2_1$ ordering, $S_{L21}$, can be defined by the following relation:

\begin{equation}\label{equ3}
\begin{split}
S_{L21} = \frac{n_{Fe}-n_{Fe}^{random}} {n_{Fe}^{full-order}-n_{Fe}^{random}},
\end{split}
\end{equation}
where n$_{Fe}^{random}$ is the number of Fe atoms on Fe sites for the random distribution,  and
n$_{Fe}^{full-order}$ is the number of Fe atoms on Fe sites in the fully ordered structure.
$S_{L21}$ can be determined from XRD measurements using the relation between
the odd superlattice and the fundamental reflections \cite{takamura09},  on the condition that  $S_{B2}$ is already known from equ.~(\ref{equ2}):

\begin{equation}\label{equ4}
\begin{split}
\frac{I_{111}^{exp}} {I_{220}^{exp}} = [S_{L21}(\frac{3-S_{B2}} {2})]^2~\frac{I_{111}^{full-order}} {I_{220}^{full-order}}.
\end{split}
\end{equation}

On this basis, we  determined the average ordering in two of our samples grown on GaAs(110) at substrate temperatures $T_{S}$ of 100 and 200~$^{\circ}$C (see Table 1).  The averaging was performed over the whole information depth of the X-rays, which is however smaller than the film thickness, and laterally over an area of several mm$^2$. The region near the FM/SC interface had to be excluded in order to avoid contributions of the GaAs lattice.  The result of 60$\%$ $L2_1$ ordering seems reasonable for thin films although in bulk material by means of long-term tempering at high temperatures, a higher degree of order is within reach. Such annealing is not possible for nanometer thick films because of accompanying interdiffusion processes, which lead to non-stoichiometry near the FM/SC interface.  On the other hand, the growth at higher substrate temperatures lead to an increased order in the as-grown film. In Ref.~\onlinecite{takamura09} the possibility of a higher  $L2_1$ ordering compared to the B2 ordering is discussed. Only a disordering of one half of the Fe and Si atoms was required for $S_{B2}$~=~0. The remaining atoms can exhibit a finite  $L2_1$-order ($S_{L21}$~$\geq$~0).

\begin{figure}[!t]
\includegraphics[width=9.0cm]{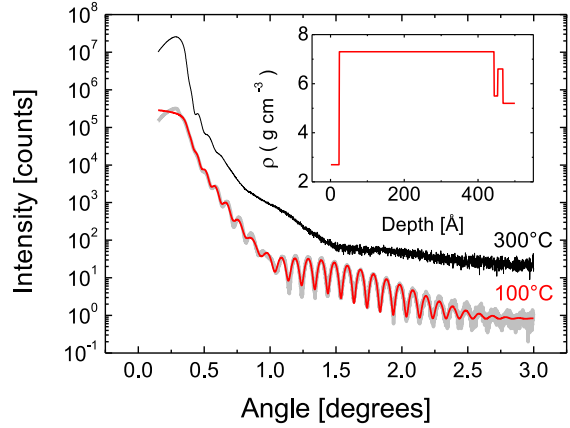}
\caption{(color online) Reflectivity curves of Co$_{2}$FeSi films of a nominal thickness of 40~nm grown on  GaAs(110) at two different substrate
temperatures T$_S$~= ~100$^\circ$C (lower curves) and  $T_{S}$~= ~300$^\circ$C (upper curve). The high quality of the surface and the FM/SC interface is shown for $T_{S}$ of 100~$^{\circ}$C by the strong interference fringes in the reflectivity curve (thicker line), which are correctly reproduced by the corresponding simulation (thin line). The inset shows the depth profile of the mass density of the Co$_{2}$FeSi film and the transition layers used in the reflectivity simulation. The reduced density near the surface is due to the native oxide. The upper reflectivity curve exhibits reduced interference fringes due to interdiffusion and/or degradation of the interface quality.
}
\label{fig:reflectivity}
\end{figure}

Figure~\ref{fig:reflectivity} demonstrates the XRR curves of two Co$_{2}$FeSi films grown on GaAs(110) at two different substrate temperatures $T_{S}$ of 100 and 300~$^{\circ}$C. The high quality of the surface and the FM/SC interface was shown for $T_S$~=~100~$^{\circ}$C by the strong interference fringes in the reflectivity curve, which were correctly reproduced by the corresponding simulation (thin line). The inset
shows the depth profile of the mass density $\varrho$ of the Co$_{2}$FeSi film and the interface region used in the reflectivity simulation. In addition an interface roughness (RMS) of $\sigma~\simeq$~0.5~nm was applied, which effectively leads to a smoothing of the density profile. The measurement results for substrate temperatures up to 250~$^{\circ}$C (not shown here) are similar, but the reflectivity curve for $T_{S}$~=~300~$^{\circ}$C already showed more pronounced differences. Interference fringes vanished nearly completely. This points to rougher interfaces and/or strong interdiffusion. Reflectivity measurements are sensitive to mass densities and not to the crystallinity and the deformation fields of the film. A characteristic mass density profile, as shown in the inset of Fig.~\ref{fig:reflectivity}, can occur for a diffusion zone \cite{geguzin1977} because diffusion is mutual: On the one hand Co, Fe, and Si diffuse into the GaAs whereas on the other hand Ga and As diffuse into the Co$_{2}$FeSi. The diffusion of Co, Fe, and Si into the GaAs was confirmed earlier by SIMS revealing that the diffusion of Co is more pronounced than the diffusion of Fe and Si.~\cite{Hentschel2012,tuck1988}

\begin{figure}[!t]
\includegraphics[width=8.0cm]{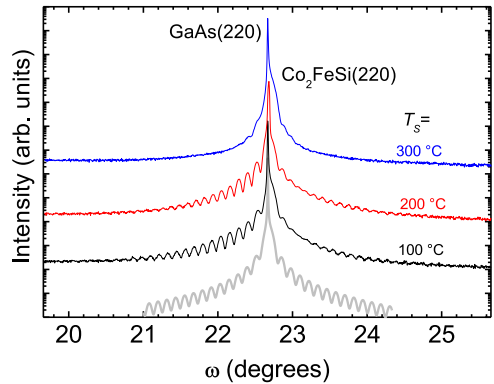}
\caption{(color online) Symmetrical X-ray diffraction peaks (220 reflections) of Co$_{2}$FeSi films of a nominal thickness of 40~nm grown on  GaAs(110) at different substrate temperatures. The lowest curve is a dynamical simulation of the diffraction pattern of a homogeneous Co$_{2}$FeSi film on GaAs. }
\label{fig:fundamental}
\end{figure}

\begin{figure}[!t]
\includegraphics[width=9.0cm]{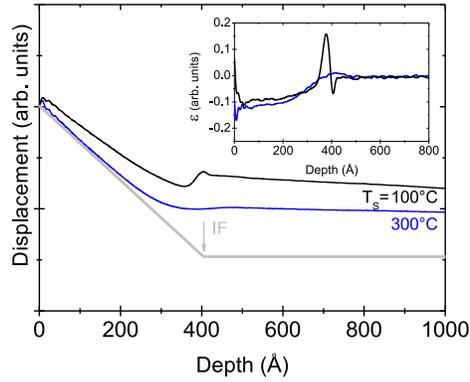}
\caption{(color online) Depth profiles of the displacement fields directly determined from two of the diffraction curves shown in Fig.~\ref{fig:fundamental} for  $T_{S}$~= ~100$^\circ$C and  $T_{S}$~= ~300$^\circ$C. The thick gray line is illustrating the idealized displacement profile for a film with homogenous lattice parameter. The inset shows corresponding deformation profiles [$\varepsilon$~(Depth)]. The arrow marks the position of the interface. Strong oscillations near the surface and the interface are visible in the inset. These oscillations are artifacts connected with Fourier transformation applied during the phase retrieval method. }
\label{fig:displacement}
\end{figure}

A symmetrical reflection ( in our case e.g. the fundamental reflection 220) can be used for a basic XRD characterization of a
heteroepitaxial film.
Figure~\ref{fig:fundamental} displays such coplanar XRD patterns of Co$_{2}$FeSi films grown on GaAs(110)  at three different substrate temperatures. The peak positions of Co$_{2}$FeSi and GaAs coincided because both materials have nearly the same lattice parameter, i.e. there was a vanishing misfit strain, the films were stoichiometric on average. The Co$_{2}$FeSi peak was broader than the GaAs substrate reflection due to the small film thickness. Interference fringes were pronounced for $T_{S}$~=~100~and~200~$^{\circ}$C indicating a high interface quality, whereas the fringes were far less visible for $T_{S}$~=~300~$^{\circ}$C. However, after more careful inspection we observed a varying amplitude of the thickness fringes similar to a beating effect. The lowest curve is a dynamical simulation of the diffraction pattern for a homogeneous Co$_{2}$FeSi film on GaAs (i.e. in the absence of any diffusion) showing interference fringes of almost constant amplitude on the logarithmic scale. Obviously the simulated diffraction pattern did not coincide with any of the experimentally observed curves.
We conjecture, that the lattice parameter changes along the direction perpendicular to the FM/SC interface. We assume, in first approximation, an inhomogeneity along the direction perpendicular to the Co$_{2}$FeSi/GaAs interface, i.e. we average laterally over an area of about one mm$^2$ as in the real X-ray experiment. We applied the method of phase retrieval X-ray diffractometry \cite{nikulin1998} for determination of the displacement depth profiles corresponding to the different diffraction curves. Figure~\ref{fig:displacement} shows the resulting depth profiles of the laterally averaged displacement field corresponding to samples grown at two different substrate temperatures $T_S$ during MBE growth. The substrate lattice was used as a reference.  Any lattice parameter different from that of the substrate introduces a displacement. The thicker gray line is illustrating the idealized displacement profile for a film with homogenous lattice parameter differing from that of the substrate. The idealized film exhibits a linear growth of the displacement with increasing distance from the interface (IF).  The real samples show a nonlinear dependence of the depth profile of displacement, i.e. the lattice parameter changes with depth [see deformation profiles $\varepsilon$(depth) shown in the inset of Fig.~\ref{fig:displacement} ]. This nonlinear depth dependence of the lattice parameter of the film leads to the fading of the interference fringes at certain diffraction angles. It may have occurred as a result of diffusion processes.   The profiles of the lattice deformations were smoother than the profile of the mass density obtained from reflectivity measurements (Fig.~\ref{fig:reflectivity}), indicating that measurable lattice deformations can be induced already by relatively low concentrations of foreign atoms thanks to the high strain sensitivity of XRD.

 From the analysis of  the XRD results, we obtained depth profiles of the lattice parameter revealing inhomogeneities inside the Co$_{2}$FeSi film.  Probably there was not only diffusion of Co, Fe, and Si into the GaAs buffer layer below the FM/SC interface but also into the opposite direction towards a native oxide and the free surface. Ga and As also diffused into the Co$_{2}$FeSi film.~\cite{geguzin1977,tuck1988,Gusenbauer2011} On the other hand direct measurements of surface temperature during MBE growth revealed an unintentional increase of $T_{S}$ during growth of about 50~K - 60~K at these low growth temperatures, which probably had an additional influence on diffusion. We saw an impact of diffusion on the structural properties of the FM/SC hybrid structures. Therefore diffusion barriers are urgently needed in order to maintain the stoichiometry and the long-range order all over the film.

\begin{figure}[!t]
\includegraphics[width=6.0cm]{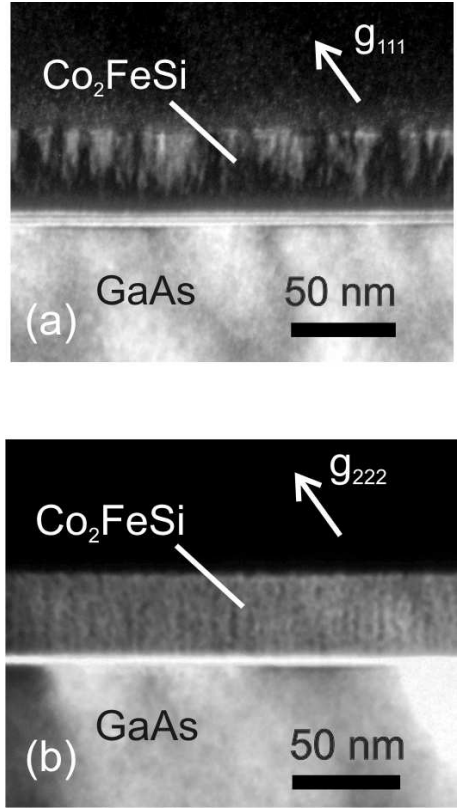}
\caption{Dark-field TEM micrographs of superlattice reflections 111 (a) and 222 (b) grown on a (110)-oriented substrate at a temperature $T_{S}$~= ~100$^\circ$C. In (a) a region of low intensity is visible near the interface and the local intensity maxima reveal well $L2_1$ ordered regions. In (b) the intensity is distributed more homogeneously although a columnar structure is visible.}
\label{fig:TEMsuper}
\end{figure}

\begin{figure}[!t]
\includegraphics[width=9.0cm]{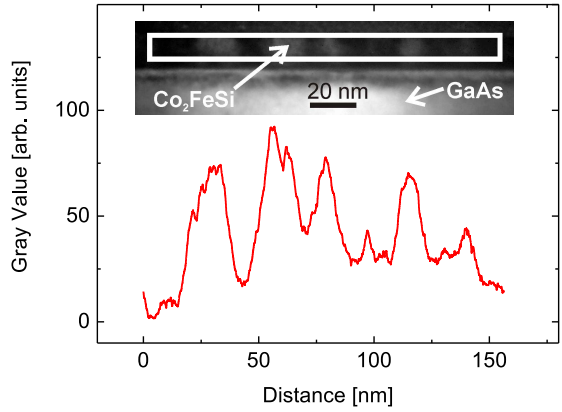}
\caption{(color online) Dark-field TEM micrograph of superlattice reflection 111 (inset) and corresponding gray values of the marked area (plot). The epitaxial Co$_{2}$FeSi thin film was grown on GaAs(111)B at a substrate temperature $T_{S}$~= ~275$^\circ$C. The inhomogeneities of the diffracted intensity
arise on roughly the same scale as the normalized Fe/Co ratio shown in the inset of Fig.~\ref{fig:eels}.}
\label{fig:gray}
\end{figure}

The average stoichiometry was tuned by minimization of the lattice mismatch between the growing Co$_{2}$FeSi film and the GaAs substrate.~ \cite{hashimoto05} However, the constituents of films and substrate have different diffusion coefficients and the amounts of atoms leaving their original positions were not equal. In this manner diffusion processes during and after growth lead to local deviations of stoichiometry. The stoichiometry itself is closely connected to the long-range ordering of the Co$_{2}$FeSi film~\cite{Jenichen05}, because only a stoichiometric material can be fully ordered.
Any nonstoichiometric alloy tends to  decompose into a mixture of ordered stoichiometric components.~\cite{khachaturyan2008,kobayashi2004} However at first the inhomogeneities of stoichiometry lead to regions of reduced crystallographic order. Figure~\ref{fig:TEMsuper} displays dark-field TEM micrographs of superlattice reflections 111 and 222 of a sample grown on a GaAs(110) substrate at a temperature $T_{S}$~= ~100$^\circ$C. In Fig.~\ref{fig:TEMsuper}~(a) we see a grainy intensity distribution (grain size $\approx$~10~nm) together with a lowering of the average intensity of the 111 reflection towards the FM/SC interface. The local intensity of the 111 reflection is a measure of the local $L2_1$ order. This $L2_1$ order exhibits a pronounced lateral inhomogeneity with the tendency of diminishing nearer to the interface. A possible explanation for a reduction of the long-range order in the vicinity of the interface would be a local nonstoichiometry caused by the predominant diffusion of Co atoms out of the Co$_{2}$FeSi film into the GaAs buffer layer. The intensity of the 222 reflection [Fig.~\ref{fig:TEMsuper}~(b)], i.e. the amount of $B2$ order was distributed more homogeneously. In our TEM experiment the scattering factors of Co and Fe were similar, and we probably do not distinguish between Co and Fe while looking at the intensity distribution of the 222 reflection. In this way, we detected mainly the disorder in the Si sublattice in this dark-field image. Figure~\ref{fig:TEMsuper}~(b) shows a columnar structure causing mainly lateral variations of the diffracted intensity similar to  the self-organized Ge$_{1-x}$Mn$_{x}$ nanocolumns in a Ge matrix~\cite{deviller2007,Yu2010}, a result of a two-dimensional spinodal decomposition.~\cite{fukushima2006} In our case such a decomposition was probably expedited by the formation of magnetic domains \cite{Felser2005,Hubert1998} during epitaxial growth of the Co$_{2}$FeSi film, which is ferromagnetic even at growth temperature. Spinodal decomposition was  already observed for other Heusler alloys.~\cite{kobayashi2004,kobayashi2006}
We found inhomogeneities of superlattice reflections also for epitaxial Co$_{2}$FeSi thin films grown on GaAs(111)B.~\cite{Jenichen2010}
In the inset of Fig.~\ref{fig:gray}, a dark-field TEM micrograph of the superlattice reflection 111 is shown as an example. The corresponding gray values of the marked area are plotted in the main part of Fig.~\ref{fig:gray}. The film was grown on GaAs(111)B at a substrate temperature $T_{S}$~= ~275$^\circ$C. On the same sample, we performed EELS measurements in the TEM.

\begin{figure}[!t]
\includegraphics[width=9.0cm]{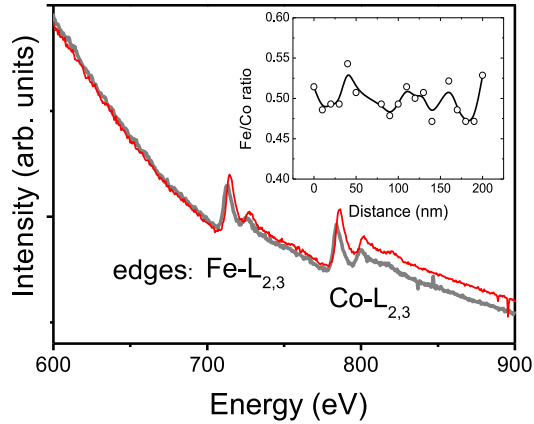}
\caption{(color online) Two EELS spectra of an epitaxial Co$_{2}$FeSi thin film on a GaAs(111)B substrate obtained in the TEM with a spatial resolution of 10~nm for two neighboring points on the sample. The Fe-L$_{2,3}$ and Co-L$_{2,3}$ edges are visible. The inset shows the spatial inhomogeneities of the normalized Fe/Co ratio obtained by EELS. The epitaxial Co$_{2}$FeSi thin film was grown on GaAs(111)B at a substrate temperature $T_{S}$~= ~275$^\circ$C.}
\label{fig:eels}
\end{figure}

EELS in the TEM provides information about the film composition with high lateral resolution. We  determined
lateral inhomogeneities of the Fe/Co composition ratio. In order to check composition inhomogeneities of Co$_{2}$FeSi we performed the measurements at many positions along a line ("line-scan") of the film cross section of the Co$_{2}$FeSi/GaAs(111)B hybrid structure (inset of Fig.~\ref{fig:eels}). Lateral inhomogeneities were revealed with relative deviations from the average composition ratio up to 6$\pm$2$\%$. The composition ratio fluctuates on a similar length scale like the intensity of the Co$_{2}$FeSi 111 superlattice reflection shown in Fig.~\ref{fig:gray}.  This finding indicates an influence of inhomogeneities of stoichiometry on the local ordering of the lattice.

\begin{figure}[!t]
\includegraphics[width=7.0cm]{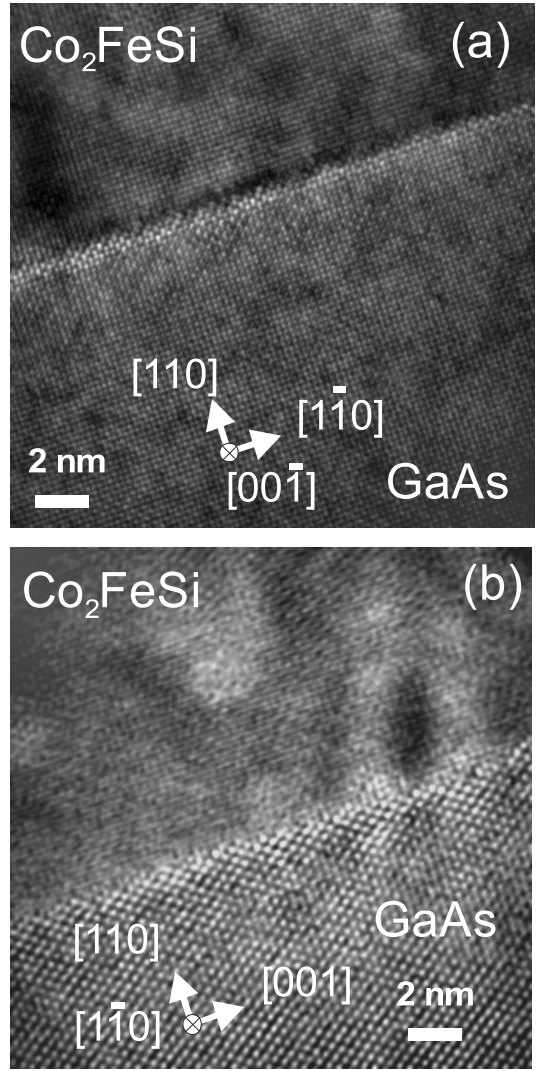}
\caption{High-resolution TEM micrographs of the interface region of a Co$_{2}$FeSi/GaAs(110) hybrid structure grown at a substrate temperature $T_{S}$~= ~100$^\circ$C. In (a) the incident beam is parallel to the GaAs [00$\bar{1}$] direction. We see a modification of the interference contrast on the GaAs side towards the interface. Inhomogeneities of the Co$_{2}$FeSi film are visible. In (b) incident beam is parallel to the GaAs [1$\bar{1}$0] direction. We again see a modification of the interference contrast on the GaAs side towards the interface. Inhomogeneities of the Co$_{2}$FeSi film appear.}
\label{fig:TEMhr}
\end{figure}

\begin{figure}[!t]
\includegraphics[width=7.0cm]{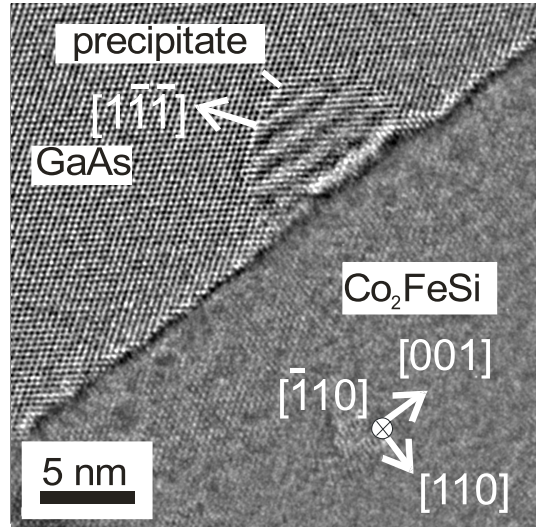}
\caption{High-resolution TEM micrograph of the interface region of a Co$_{2}$FeSi/GaAs(110) hybrid structure grown at a substrate temperature $T_{S}$~= ~200$^\circ$C. The beginning stage of the formation of a precipitate becomes visible by Moire contrast. A [1$\bar{1}\bar{1}$] facet is already observed. }
\label{fig:TEMprecipitate}
\end{figure}

Often HR TEM micrographs are taken as evidence for the high structural quality of Heusler alloy films. However in this mode of operation of the TEM many reflections interfere. The superlattice reflections were of low intensity compared to the fundamental ones. One has to pay attention to minor changes in the interference contrast in order to observe effects due to interdiffusion and/or ordering. Figure~\ref{fig:TEMhr} (a) shows a high-resolution TEM micrograph of the FM/SC interface region of a Co$_{2}$FeSi/GaAs(110) hybrid structure grown at a substrate temperature $T_{S}$~= ~100$^\circ$C.  The incident beam was parallel to GaAs [00$\bar{1}$]. A high interface quality was found, however looking more carefully we see a modification of the GaAs interference contrast towards the interface, probably connected to interdiffusion. Inhomogeneities of the Co$_{2}$FeSi film occurred and the contrast distribution of the Co$_{2}$FeSi near the interface is modified as well. Figure~\ref{fig:TEMhr} (b) shows a high-resolution TEM micrograph of the same structure as Fig.~\ref{fig:TEMhr}(a) cut along a perpendicular plane, i.e. the incident electron beam is now parallel to GaAs [1$\bar{1}$0]. Again a modification of the GaAs contrast towards the interface can be observed. Inhomogeneities of the Co$_{2}$FeSi film appear as well along this projection. HR TEM has been applied earlier for characterization of Co$_{2}$FeSi grown on GaAs(001).~\cite{hashimoto2007jvst} There, the partly disordered $B2$ phase was found near the FM/SC interface. Another issue is the interface stability with respect to the formation of precipitates shown in Fig.~\ref{fig:TEMprecipitate}: In our structures precipitation near the interface occurred at $T_{S}$~= ~200$^\circ$C \cite{Hentschel2012} similar to the structures grown on GaAs(001) \cite{hashimoto2007jvst}. For Co$_{2}$FeSi/GaAs(111)B hybrid structures this critical temperature is $T_{S}$~= ~275$^\circ$C indicating an improved stability  of the 111 oriented FM/SC interface \cite{Jenichen2010} compared to the 110 and 001 orientations of the interface. However, the formation of precipitates in restricted areas of the interface may turn out less critical than expected, because the remaining areas of the interface still exhibit high perfection with a slightly increased interface roughness (see Fig.~\ref{fig:TEMprecipitate}) and the precipitates themselves may getter foreign atoms in GaAs and reduce in this way the scattering of spins.~\cite{ramsteiner08} The better alternative though is to avoid the diffusion of Co, Fe, and Si into GaAs and of Ga and As into Co$_{2}$FeSi by introduction of a diffusion barrier.

\section{Conclusions}
Co$_{2}$FeSi films  can be grown on GaAs highly but not fully ordered. Interdiffusion between Co$_{2}$FeSi and GaAs, as well as segregation or diffusion towards the free surface and/or the native oxide lead to inhomogeneous depth profiles of the lattice parameter caused by local variations of film composition. These depth profiles were superimposed on a grainy distribution of the long-range order of the Co$_{2}$FeSi lattice, which was probably connected to the formation of magnetic domains. Local deviations from stoichiometry reduced the ordering inside the Heusler alloy and lead to decomposition. Such inhomogeneities of the long-range order were found for both substrate orientations GaAs(111)B and GaAs(110). The thermal stability of the 111--interface is higher, i.e. precipitation near the interface was found only at a  substrate temperature higher by 75~K.

\section{Acknowledgement}
The authors thank Claudia Herrmann, and Hans-Peter Sch\"onherr for their support during the
MBE growth, Doreen Steffen for sample preparation, Astrid Pfeiffer
for help in the laboratory,  Esperanza Luna and Uwe Jahn
for valuable support and helpful discussion. AYN acknowledges support of the ARC Centre of Excellence in Coherent X-Ray Science.

\section{References}

%

\newpage

\end{document}